\documentclass{article}
\usepackage{spconf,amsmath,graphicx,verbatim,amsfonts,xcolor,bm,multirow,subfigure}


\title{Incremental Binarization On Recurrent Neural Networks\\For Single-Channel Source Separation}

\name{Sunwoo Kim, Mrinmoy Maity, Minje Kim\thanks{This project was supported by Intel Corporation.}}
\address{Indiana University\\
    Department of Intelligent Systems Engineering\\
    Bloomington, IN 47408\\
    {\small \texttt{kimsunw@indiana.edu},~ \texttt{mmaity@iu.edu}, ~\texttt{minje@indiana.edu}}
}

\begin{document}
\ninept

\maketitle

\begin{abstract}
This paper proposes a Bitwise Gated Recurrent Unit (BGRU) network for the single-channel source separation task. Recurrent Neural Networks (RNN) require several sets of weights within its cells, which significantly increases the computational cost compared to the fully-connected networks. To mitigate this increased computation, we focus on the GRU cells and quantize the feedforward procedure with binarized values and bitwise operations. The BGRU network is trained in two stages. The real-valued weights are pretrained and transferred to the bitwise network, which are then incrementally binarized to minimize the potential loss that can occur from a sudden introduction of quantization. As the proposed binarization technique turns only a few randomly chosen parameters into their binary versions, it gives the network training procedure a chance to gently adapt to the partly quantized version of the network. It eventually achieves the full binarization by incrementally increasing the amount of binarization over the iterations. Our experiments show that the proposed BGRU method produces source separation results greater than that of a real-valued fully connected network, with 11-12 dB mean Signal-to-Distortion Ratio (SDR). A fully binarized BGRU still outperforms a Bitwise Neural Network (BNN) by 1-2 dB even with less number of layers.

\end{abstract}


\section{Introduction}
Neural network-based approaches to source separation tasks have been becoming more prevalent \cite{xu2014experimental,huang2015joint, nugraha2016multichannel}. Fully connected deep neural networks (DNN) have shown to be capable of learning complex mapping functions from a large set of noisy signals and their corresponding ideal binary mask (IBM) target outputs \cite{wang2013towards,le2015deep,grais2014deep}. Recurrent neural networks (RNN), which are structured to be more effective in applications involving sequential or temporal data, have also shown to excel in the same task \cite{erdogan2015phase,weninger2015speech,weninger2014discriminatively,isik2016single,chen2015speech}. The RNN is able to attain the superior performance by utilizing a shared hidden state and gates within its hidden cells that guide the memory and learning over a sequence of inputs \cite{bengio1994learning}. The most practical method to train RNNs is with truncated Backpropagation Through Time (BPTT) \cite{williams1990efficient}. This bounded-history approximation method simplifies computation by limiting itself to a fixed scope of $T$ timesteps \cite{sutskever2013training}. 

The most efficient cell structure that is robust to the gradient vanishing problem is the Gated Recurrent Unit (GRU) cell \cite{cho2014learning}. The computation within each GRU cell is:
\begin{equation} \label{eq:gru}
    \begin{split}
        \mathbf{r}^{(l)}(t) &= \sigma\Big(\mathbf{W}^{(l)}_r\mathbf{x}^{(l-1)}(t) + \mathbf{U}^{(l)}_r\mathbf{h}^{(l)}(t-1)\Big)\\
        \mathbf{z}^{(l)}(t) &= \sigma\Big(\mathbf{W}^{(l)}_z\mathbf{x}^{(l-1)}(t) + \mathbf{U}^{(l)}_z\mathbf{h}(t-1)\Big)\\
        \tilde{\mathbf{h}}^{(l)}(t) &= \phi\Big(\mathbf{W}^{(l)}_h\mathbf{x}^{(l-1)}(t) + \mathbf{U}^{(l)}_h \big(\mathbf{r}^{(l)}(t)\odot\mathbf{h}^{(l)}(t-1)\big)\Big)\\
        \mathbf{h}^{(l)}(t) &= \mathbf{z}^{(l)}(t)\odot \mathbf{h}^{(l)}(t-1) + (1-\mathbf{z}^{(l)}(t)) \odot \tilde{\mathbf{h}}^{(l)}(t)\\
    \end{split}
\end{equation}
where $l=\{1,\dots,L+1\}$ denotes the layer index and $t=\{1,\dots,T\}$ is the time index. $\mathbf{r}_t, \mathbf{z}_t,\tilde{\mathbf{h}}_t,$ and $\mathbf{h}_t$ are reset gate, update gate, candidate hidden state, and updated hidden state respectively all of dimension $\mathbb{R}^{K^{(l)}}$ with $K^{(l)}$ as the number of units at layer $l$. $\mathbf{W}_r^{(l)}\in \mathbb{R}^{K^{(l)}\times K^{(l-1)}}$ and $\mathbf{U}^{(l)}_r\in \mathbb{R}^{K^{(l)}\times K^{(l)}}$ are the weight matrices for the input $\mathbf{x}^{(l)}(t)$ and previous hidden state $\mathbf{h}^{(l)}(t-1)$ at the reset gate. Similarly, $\mathbf{W}_z$, $\mathbf{U}_z$, $\mathbf{W}_h$, and $\mathbf{U}_h$ are corresponding weights for the update gate and candidate state. The $\sigma$ and $\phi$ refer to the logistic sigmoid and hyperbolic tangent activation functions. The bias term is omitted for simplicity. Note that $\mathbf{h}^{(l)}(t)$ is fed to the next layer as an input, $\mathbf{x}^{(l)}(t)$.

For a single feedforward step, the RNN requires multiple sets of weights and performs operations in (\ref{eq:gru}) for $T$ timesteps. With deeper RNNs, the computational cost rises rapidly in terms of $K$ and $L$. This paper presents an efficient method to reduce the computational and spatial complexity of the GRU network for the source separation problem while maintaining high performance results. We extend from the idea of Bitwise Neural Networks (BNN)  \cite{kim2018bitwise} \cite{courbariaux2015binaryconnect} and low-precision RNNs \cite{ott2016recurrent}. The model we propose is a Bitwise GRU (BGRU) network that reduces network complexity by re-defining the originally real-valued inputs and outputs, weights, and operations in a bitwise fashion. By limiting the network to bipolar binary values, the space complexity of the network can be significantly reduced. In addition, all real-valued operations during the feedforward procedure can be replaced with bitwise logic, which further reduces both spatial and time complexity \cite{hubara2016binarized,rastegari2016xnor,govindu2004analysis,beauchamp2006embedded}. 

Transforming real-valued weights into bipolar binaries results in heavy quantization loss \cite{hwang2014fixed,courbariaux2014training}. To alleviate this effect, the weights are converted into binary values through a gentle training procedure. In this paper, we introduce an incremental training method for weights of the BGRU network that holds onto the quality of the source separation model. Experimental results for single-channel source separation tasks show that the BGRU model shows incremental and predictable loss depending on the amount of binarization and still performs better than a real-valued Fully-Connected Network (FCN).  

\section{Bitwise Gated Recurrent Units (BGRU)}

\subsection{Background: Bitwise Neural Networks}

Binarization has been explored as a method of network compression. BinaryConnect \cite{courbariaux2015binaryconnect}, binarized neural networks \cite{hubara2016binarized}, trained ternary quantization \cite{ZhuC2016arxiv}, and Bitwise Neural Networks (BNN) \cite{kim2015bitwise} have implemented a binarized or ternarized neural network in bipolar binaries (with zeros in the ternarized case) for network compression. They emphasize that replacing real-valued operations with bitwise versions greatly reduces the network's complexity. In particular, the BNN training process is assisted by initializing the binarized network with pretrained weights. The weights are compressed in a real-valued network with the hyperbolic tangent activation function in order to better approximate their binary versions. Further quantization is performed in the BNN, where the inputs are quantized using Quantization-and-Disperson, which uses Lloyd-Max's quantization to convert each frequency magnitude of the noisy input spectrum into 4 bits with bipolar binary features \cite{lloyd1982least}. In the domain of source separation, BNN's have been applied by predicting Ideal Binary Masks (IBM) as target outputs \cite{kim2018bitwise}.  

While the BNN significantly reduces the space and time complexity of the network, the conversion from real-values to bipolar binaries inevitably produces quantization error. One method to reduce this penalty is the concept of sparsity \cite{kim2018bitwise}. Sparsity can be introduced to bitwise networks by converting the pretrained weights with smaller values to 0's. The threshold for determining the sparsity is calculated with a predefined boundary $\beta$. The relaxed quantization process for a weight element $w$ is:  
\begin{equation} \label{eq:sparsity}
    \bar{w}=\begin{cases}
               +1 \;\;\text{if}\;\; w > \beta\\
               -1 \;\;\text{if}\;\; w < -\beta\\
                0 \;\;\;\;\:\text{otherwise}
            \end{cases}
\end{equation}
where $\bar{w}$ represents the binarized variable. Another way to mitigate the quantization error is by multiplying a scaling factor $\mu$ to the bipolar-binarized weights, so that the quantized values approximate the original values more closely \cite{ZhuC2016arxiv}. 


\subsection{Feedforward in BGRU}

\subsubsection{Notation and setup}

For the following sections of the paper, we specify discrete variables with a bar notation, i.e. $\bar{x}$. Depending on the context, this could be a binary variable with 0 and 1 (e.g. gates), a bipolar binary variable with $+1$ and $-1$ (e.g. binarized hidden units), or a ternary variable (e.g. sparse bipolar binary weights). The binary versions of logistic sigmoid and hyperbolic tangent activation functions are:
\begin{equation}\label{eq:bsigmoid}
\bar{\sigma}(x)= \frac{sgn(x)+1}{2} \in\{0,1\},  ~\bar{\phi}(x) = sgn(x) \in\{-1,+1\}
\end{equation}
respectively where $sgn(x)$ is a sign function \cite{hubara2016binarized, kim2015bitwise}. 

\subsubsection{The feedforward procedure}
In the BGRU, the feedforward process is defined as follows:
\begin{equation} \label{eq:bgru}
    \begin{split}
        \bar{\mathbf{r}}^{(l)}(t) &= \bar{\sigma}\Big(\bar{\mathbf{W}}^{(l)}_r \bar{\mathbf{x}}^{(l-1)}(t) + \bar{\mathbf{U}}^{(l)}_r\bar{\mathbf{h}}^{(l)}(t-1)\Big)\\
        \bar{\mathbf{z}}^{(l)}(t) &= \bar{\sigma}\Big(\bar{\mathbf{W}}^{(l)}_z \bar{\mathbf{x}}^{(l-1)}(t) + \bar{\mathbf{U}}^{(l)}_z\bar{\mathbf{h}}^{(l)}(t-1)\Big)\\
        \bar{\tilde{\mathbf{h}}}^{(l)}(t) &= \bar{\phi}\Big(\bar{\mathbf{W}}^{(l)}_{h}\bar{\mathbf{x}}^{(l-1)}(t) +\bar{\mathbf{U}}^{(l)}_h \big(\bar{\mathbf{r}}^{(l)}(t)\odot\bar{\mathbf{h}}^{(l)}(t-1)\big)\Big)\\
        \bar{\mathbf{h}}^{(l)}(t) &= \bar{\mathbf{z}}^{(l)}(t)\odot\bar{\mathbf{h}}^{(l)}(t-1) + (1-\bar{\mathbf{z}}^{(l)}(t))\odot\bar{\tilde{\mathbf{h}}}^{(l)}(t)\\
    \end{split}
\end{equation}
The product between two binarized values (e.g. between the $(i,j)$-th element of $\bar{\mathbf{W}}^{(l)}_r$ and the $j$-th element of $\bar{\mathbf{x}}^{(l-1)}(t)$) is equivalent to the XNOR operations, a cheaper binary operation than the corresponding floating-point operation. Also, the use of sign functions $\bar{\phi}$ and a hard step function $\bar{\sigma}$ in place of the hyperbolic tangent and sigmoid functions also expedite the process because they can be usually implemented by a pop counter. 

\subsubsection{Scaled sparsity and Bernoulli masks}

We define two types of masks that are applied on various parts of the network. The scaled sparsity mask is a two-in-one solution to introduce both scaling parameters and sparsity into weights during the binarization process. To binarize the weight matrices the scaled sparsity mask $\mathbf{B}$ is created using a predefined sparsity parameter $0\!<\!\rho\!<\!1$. First, we find a per-layer cutoff value $\beta$ and the scaling parameter $\mu$ that meet the following equations: 
\begin{equation}\label{eq:beta}
    \begin{split}
        \mathcal{S} &=\{(i,j): |W^{(l)}_{i,j}|>\beta\}, \quad |\mathcal{S}| = K^{(l-1)}K^{(l)}\rho\\
        \mu &= \frac{1}{|\mathcal{S}|}\sum_{(i,j)\in\mathcal{S}} |W^{(l)}_{i,j}|,
    \end{split}
\end{equation}
where $\mathcal{S}$ is the set of weight indices whose absolute values are larger than the cutoff value and $|\mathcal{S}|$ denotes the number of such weights. Therefore, for a given sparsity value $\rho$, we first sort the weights in their absolute values and then find the cutoff that results in $\mathcal{S}$ with the predefined size. Using $\beta$ and $\mu$, we set the mask elements as follows:
\begin{equation} \label{eq:msmask}
    B_{i,j}=\begin{cases}
               \mu \;\;\text{if}\;\; |W_{i,j}| > \beta\\
                0 \;\;\;\;\:\text{otherwise}
            \end{cases}
\end{equation}

The other type of mask is a random Bernoulli matrix $\mathbf{C}$ with a parameter $0<\pi<1$ as the amount of binarization. The value of $\pi$ is initially chosen as a small value (e.g. 0.1 for 10\% binarization) and gradually increased up to 1.0, which means the network is completely binarized. The created masks are applied on weights $\mathbf{W}$ to create the partly binarized matrix $\widehat{\mathbf{W}}$: 
\begin{equation}\label{eq:sparsity_mask}
  \widehat{\mathbf{W}} = \big(\bar{\phi}(\mathbf{W})\odot\mathbf{B}\big)\odot\mathbf{C} + \phi(\mathbf{W})\odot(1-\mathbf{C}).  
\end{equation} 

The purpose of $\mathbf{B}$ with $\mu$ values is to lessen the quantization error from the binarization. The $\bar{\phi}$ operator will transform all values into bipolar binary values, which would be too intensive of a tranformation because the distribution of the first round weights are all relatively close to 0. Thus, by multiplying the remaining nonzero bipolar values after applying sparsity with $\mu$, the values are scaled down to the average value of the non-sparse portion, which is a better representative for the nonzero elements. Note that feedforward is still bitwise thanks to the symmetry of $\mathbf{B}$ and by skipping zeros. 

The Bernoulli mask $\mathbf{C}$ enables a gradual transition from real-valued weights and operations to bitwise versions. This mask is applied on the bitwise and real-valued elements in a complementary way to control the proportion of binarization in the network. $\widehat{\mathbf{W}}$ in (\ref{eq:sparsity_mask}) is binarized only partly with the proportion set by $\pi$. 
Note that for the real-valued weights we are using a $\tanh$ compressed version $\phi(\mathbf{W})$ for the purpose of regularization (see Section \ref{sec:1stR} for more details). 

$\mathbf{C}$ is used to control the binarization of the other network elements such as gates and hidden units, too. For the candidate hidden units $\bar{\tilde{\mathbf{h}}}$, for example, the activations are performed as:
\begin{equation}\label{eq:bernoulli_mask}
  \widehat{\tilde{\mathbf{h}}} = \bar{\tilde{\mathbf{h}}}\odot\mathbf{C} + \mathbf{\tilde{\mathbf{h}}}\odot(1-\mathbf{C}),  
\end{equation}
The gates are also partially binarized in this way. The mask $\mathbf{C}$ is generated at each iteration for the weights as in (\ref{eq:sparsity_mask}) and then at each timestep for the activation functions of GRU cells as in (\ref{eq:bernoulli_mask}). This ensures that the gradients of the bitwise terms are evenly distributed gradually for all weights at each levels of $\pi$. 
Without even distribution, certain elements of the graph that do not participate in the bitwise procedure begin focusing on compensating for the quantization loss from the other bitwise elements. This needs to be avoided since as $\pi$ is increased to 1.0 these elements need to be quantized eventually. 

\subsection{Training BGRU Networks}
The objective is to accept binarized mixture signal inputs and predict the corresponding IBMs. The inputs are binarized using the Quantization-and-Dispersion technique \cite{kim2015bitwise}, and the target outputs are bipolar binary IBM predictions which are later converted to 0's and 1's for the source separation task. 
We follow the typical two-round training scheme from BNNs, too. 

\subsubsection{First round: Pretraining $\phi$-compressed weights}\label{sec:1stR}
The GRU network is first initialized with real-valued weights and then trained on quantized binary inputs. During training, the weights are wrapped with the hyperbolic tangent activation function, $\phi$. This has the effect of bounding the range of weights between $-1$ and $+1$ as well as regularization. In the second round, the sign function, $\bar{\phi}$ is applied on the weights instead, hence the first round network can be perceived as its softer version. For example, the feedforward procedure in \eqref{eq:gru} for only the hidden candidate state at layer $l$ and timestep $t$ becomes:
\begin{equation}\label{eq:compressed_rt}
    \tilde{\mathbf{h}}^{(l)}\!(t)\!=\!\phi\Big(\!\phi\big(\mathbf{W}^{(l)}_h\big) \mathbf{x}^{(l-1)}\!(t) + \phi\big(\mathbf{U}_h^{(l)}\big) \big(\bar{\mathbf{r}}^{(l)}(t)\odot\bar{\mathbf{h}}^{(l)}\!(t\!-\!1)\big)\!\Big)\!\!\!
\end{equation}
The $\phi$-compressed weights are applied similarly for the reset and update gates. 

\textit{\textbf{Backpropagation}}: With the introduction of $\phi$ on the weight matrices, the derivative with respect to $\phi$ is added onto the backpropagation due to the chain rule. For example, the gradients for \eqref{eq:compressed_rt} are computed as:
\begin{align}\label{eq:first_gradient}
\nonumber\bm{\delta}_{\tilde{\mathbf{h}}}(t) &= \bm{\delta}^{(l)}(t) \odot (1-\mathbf{z}(t)) \odot \big(1-\tilde{\mathbf{h}}^{(l)}(t)^2\big) \\
\nabla \mathbf{W}^{(l)}_h &= \Big(\sum_{t=0}^T \bm{\delta}_{\tilde{\mathbf{h}}}(t) \cdot \big(\mathbf{x}^{(l-1)}(t)\big)^\top\Big)\odot \Big(1-\phi^2\big(\mathbf{W}^{(l)}_h\big)\Big)\\
\nonumber\nabla \mathbf{U}^{(l)}_h &= \Big(\sum_{t=1}^T \bm{\delta}_{\tilde{\mathbf{h}}}(t) \cdot \big(\bar{\mathbf{r}}^{(l)}(t)\odot\bar{\mathbf{h}}^{(l)}(t-1)\big)\Big)\odot \Big(1-\phi^2\big(\mathbf{U}^{(l)}_h\big)\Big)
\end{align}
where $\bm{\delta}^{(l)}(t)$ is the backpropagation error for the training sample at layer $l$ and timestep $t$. The gradients are similarly defined for the weights in the gates.

\subsubsection{Second round: BGRU}
The BGRU network is initialized with the real-valued weights from the first round, which are pretrained to be optimal for the source separation task. The real-valued weights are saved for the backpropagation step and used to construct bitwise weights for the feedforward procedure using both the mean-scaled sparsity mask $\mathbf{B}$ and Bernoulli mask $\mathbf{C}$. The bitwise activation functions, $\bar{\sigma}$ and $\bar{\phi}$ are applied during the feedforward as well. Again as an example, with the introduction of the masks and bitwise functions, the feedforward step for the hidden candidate state becomes: 
\begin{equation}\label{eq:second_round}
\begin{split}
    \widehat{\mathbf{W}}^{(l)}_h &= (\bar{\phi}(\mathbf{W}^{(l)}_h) \odot \mathbf{B}) \odot \mathbf{C} + \phi(\mathbf{W}^{(l)}_h) \odot (1-\mathbf{C})\\
    \widehat{\mathbf{U}}^{(l)}_h &= (\bar{\phi}(\mathbf{U}^{(l)}_h) \odot \mathbf{B}) \odot \mathbf{C} + \phi(\mathbf{U}^{(l)}_h) \odot (1-\mathbf{C})\\[2pt]
    \mathbf{V} &= \widehat{\mathbf{W}}^{(l)}_h \mathbf{x}^{(l-1)}(t) + \widehat{\mathbf{U}}_h^{(l)} \big(\bar{\mathbf{r}}^{(l)}(t)\odot\bar{\mathbf{h}}^{(l)}(t-1)\big)\\
    \widehat{\tilde{\mathbf{h}}}^{(l)}(t) &= \bar{\phi}(\mathbf{V})\odot \mathbf{C} + \phi(\mathbf{V})\odot (1-\mathbf{C})
\end{split}
\end{equation}
where $\mathbf{V}$ is an intermediary term. The Bernoulli parameter $\pi$ is incremented gradually to determine $\mathbf{C}$ until the network is completely binarized at $\pi=1.0$. 

\textbf{\textit{Backpropagation:}} The derivatives of non-differentiable activation functions are overwritten with the derivatives of their relaxed counterparts such that $\bar{\phi}'=\phi'$ and $\bar{\sigma}'=\sigma'$. This simplifies the gradients for (\ref{eq:second_round}). The gradients are computed as (\ref{eq:first_gradient}) with an additional factor for the masks which are:
\begin{equation}\label{eq:second_grad}
    \begin{split}
        \nabla \mathbf{W}^{(l)}_h &= \nabla \mathbf{W}^{(l)}_h \odot (\mathbf{B}\odot \mathbf{C} + (1-\mathbf{C}))\\
        \nabla \mathbf{U}^{(l)}_h &= \nabla \mathbf{U}^{(l)}_h \odot (\mathbf{B}\odot \mathbf{C} + (1-\mathbf{C}))\\
    \end{split}    
\end{equation}
The gradients are computed similarly for the gates. The calculations in (\ref{eq:second_grad}) show that the network is the same as the first round network except with the addition of masking factors. Only the real-valued weights are updated with the gradients during training. 

\section{Experiments}

\subsection{Experimental Setups}

For the experiment, we randomly subsample 12 speakers for training and 4 speakers for testing from the TIMIT corpus. For both subsamples, we select half of the speakers as male and the other half as female. There are 10 short utterances per speaker recorded with a 16kHz sampling rate. Each utterances are mixed with 10 different non-stationary noise signals with 0 dB Signal-to-Noise Ratio (SDR), namely $\{$birds, casino, cicadas, computer keyboard, eating chips, frogs, jungle, machine guns, motorcycles, ocean$\}$ \cite{duan2012online}. In total, we have 227,580 training examples and 81,770 test examples from 1,200 and 400 mixed utterances, respectively. We apply a Short-Time Fourier Transform (STFT) with a Hann window of 1024 and hop size of 256. To quantize the spectra into bipolar binaries, we apply a 4-bit QaD procedure and convert them into $n\times (4 \times 513)$ dimension matrices. These vectors are used as inputs to the BGRU systems. The truncated BPTT length used was $T=50$. We found $\rho=0.8$ to perform well in our experiment. We used the Adam optimizer for both first and second rounds with the same beta parameters, $\beta_1=0.4$ and $\beta_2=0.9$. Minibatch size is set as 10 for 10 mixed utterances constructed from 1 clean signal mixed with the 10 noise signals. We train two types of networks that predict the IBMs with respect to the noisy quantized input:


\begin{figure*}[t!]
    \centering
\subfigure[Results from 1000 epochs for $\pi<1.0$ and 100 epochs for $\pi=1.0$]{\includegraphics[keepaspectratio, width=\columnwidth, height=5cm]{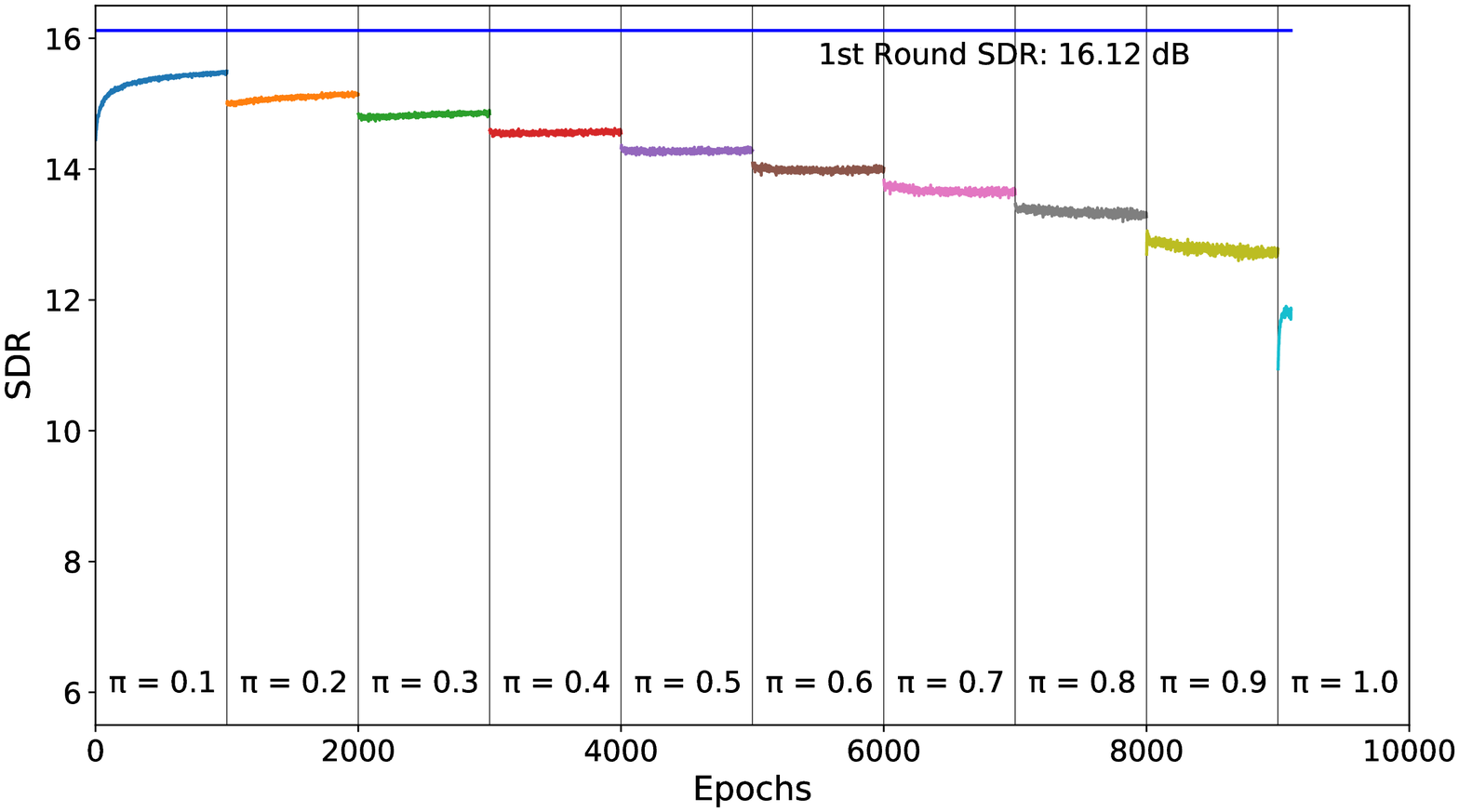}\label{fig:subfig1}}
\subfigure[Results from 100 epochs for each $\pi$]{\includegraphics[keepaspectratio, width=\columnwidth, height=4.7975cm]{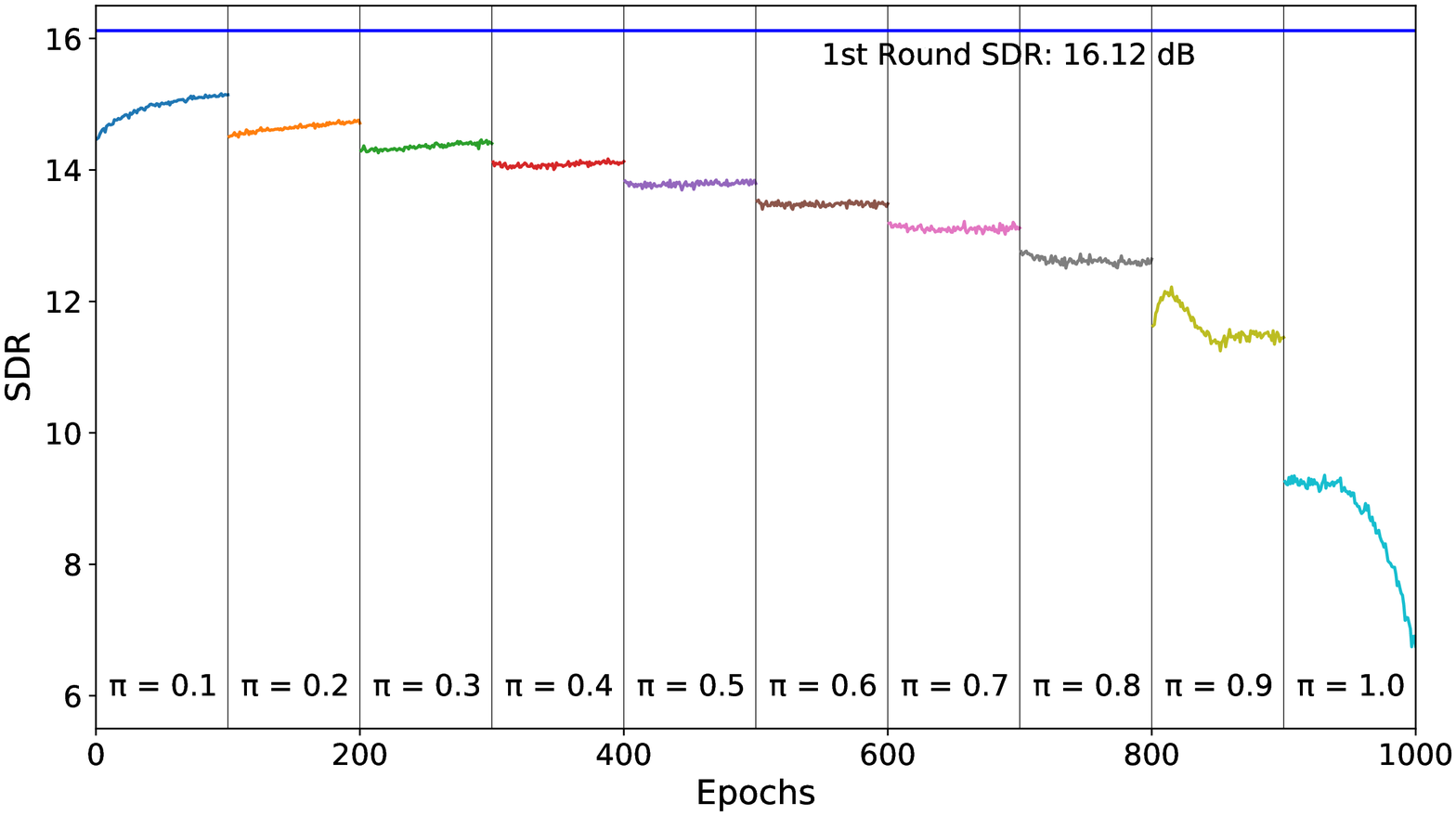}\label{fig:subfig2}}
    \caption{Second round testing results on incremental levels of $\pi$. Figures \subref{fig:subfig1} and \subref{fig:subfig2} show the effects of running different number of iterations.}
    \label{fig:inc_train}
\end{figure*}

\begin{table}[t!]
\centering
\caption{Speech denoising performance of the proposed BGRU-based source separation model compared to FCN, BNN, and GRU networks} \label{tab:table}
\vspace{1mm}
\begin{tabular}{|c|c|c|c|c|}
\hline
 \multicolumn{2}{|c|}{Systems} & Topology & SDR & STOI \\
 \hline
 \multicolumn{2}{|c|}{\multirow{2}{*}{FCN with original input}}
 & 1024$\times$2 & 10.17 & 0.7880 \\
 \multicolumn{2}{|c|}{}& 2048$\times$2 & 10.57 & 0.8060 \\
 \hline 
 \multicolumn{2}{|c|}{\multirow{2}{*}{FCN with binary input} }
 & 1024$\times$2 & 9.80 & 0.7790 \\
 \multicolumn{2}{|c|}{}& 2048$\times$2 & 10.11 & 0.7946 \\
 \hline
 \multicolumn{2}{|c|}{\multirow{2}{*}{BNN} }
 & 1024$\times$2 & 9.35 & 0.7819 \\
 \multicolumn{2}{|c|}{} & 2048$\times$2 & 9.82 & 0.7861 \\
 \hline
 \multicolumn{2}{|c|}{GRU with binary input} & 1024$\times$1 & 16.12 & 0.9459 \\
 \hline
 \multirow{10}{*}{BGRU} & $\pi$=0.1 & \multirow{10}{*}{1024$\times$1} & 15.50 & 0.9393 \\
 & $\pi$=0.2 & & 15.17 & 0.9361 \\
 & $\pi$=0.3 & & 14.90 & 0.9324 \\
 & $\pi$=0.4 & & 14.58 & 0.9292 \\
 & $\pi$=0.5 & & 14.32 & 0.9252 \\
 & $\pi$=0.6 & & 14.02 & 0.9217 \\
 & $\pi$=0.7 & & 13.66 & 0.9174 \\
 & $\pi$=0.8 & & 13.30 & 0.9104 \\
 & $\pi$=0.9 & & 12.70 & 0.9019 \\
 & $\pi$=1.0 & & 11.76 & 0.8740 \\
 \hline
\end{tabular}
\end{table}

\begin{list}{\labelitemi}{\leftmargin=1em}
\item \textit{Baseline with binary input}: The baseline network is constructed with a single GRU layer with $K=1024$ units. The inputs to the network are $4\times 513$ dimension 4-bit QaD vectors and predicted outputs are $513$ dimension IBMs. We use the first round training algorithm to train the baseline network. For regularization, we apply dropout rate of 0.05 for the input layer and 0.2 for the GRU layers.
\item \textit{The proposed BGRU}: We initialize the weights with the pretrained weights and use the second round training algorithm to train the BGRU network. We increase the $\pi$ parameter by 0.1 starting from 0.1 to 1.0. The learning rates are reduced for each increase in $\pi$.
\end{list}

\subsection{Discussion}
Table \ref{tab:table} shows results for the BGRU along with other systems for comparison. The metrics displayed are Signal-to-Distortion Ratio (SDR) \cite{vincent2006performance} and Short-Time Objective Intelligibility (STOI) \cite{taal2010short}. At each increase in $\pi$, there is a distinct drop in SDR and STOI due to the loss in information as we increase the number of elements undergoing binarization. 
Since the initial weights transferred from the first round are optimal, we restrict the weights from updating too drastically by dampening the learning rate at each increase in $\pi$. 
We did not observe substantial difference from reducing the learning rate before $\pi=0.8$, however the performance becomes sensitive as the rate of binarization nears 1. In Figure \ref{fig:subfig1} it can be seen that from $\pi=0.8$ the performance begins to decrease more than during previous $\pi$ values. 

The BGRU network is trained for an extended number of iterations so it propagates the corrections and adjusts to the quantization injected into the network.
We trained 1000 epochs for each $\pi$ values except at $\pi=1.0$. Figure \ref{fig:subfig1} shows that this many iterations is not always beneficial within the same session with a fixed $\pi$, because SDR improvement becomes stagnant and even starts to drop. However, in this way the network can prevent a greater drop in performance at the next increase in $\pi$. At $\pi=1.0$, we only train for 100 epochs and perform early stopping because the network is less robust and degrades in performance after more than 100 epochs. Also, since the network has finished training for the source separation task at $\pi=1.0$, further training is unnecessary.
On the contrary, Figure \ref{fig:subfig2} shows that training for less number of iterations, e.g. 100 epochs, produces a greater drop at each increment of $\pi$. 

The drop in performance from a real-valued network to a bitwise version is quite comparable between a FCN with BNN and GRU with BGRU. The loss is much greater in the BGRU network (16.12 dB to 11.76 dB SDR) than in the case of BNN (10.11 dB to 9.82 dB SDR). 
Yet, the performance of a single-layer fully bitwise BGRU network with 1024 units (11.76 dB SDR and 0.8740 STOI) is still greater than that of a double-layer BNN with 2048 units (9.82 dB SDR and 0.7861 STOI), and also greater than that of a unquantized double-layer FCN with real-valued inputs and 2048 units (10.57 dB SDR and 0.8060 STOI). We discuss the space complexity of the BGRU network compared to a FCN and BNN. Considering that a GRU layer contains 3 sets of weights, the single layer BGRU network contains $3\times (1024\times1)$ number of weights. This number is still less than a FCN or BNN of topology $2048\times2$. We introduced a real-valued scaling factor $\mu$, but it reduces down to bipolar binaries once training is done, so it does not add additional costs. 

In the future, we plan to extend the network structure to deeper ones. Also, more scheduled annealing of the $\pi$ values is another option to investigate.

\section{Conclusion}
In this paper, we proposed an incremental binarization procedure to binarize a RNN with GRU cells. The training is done in two rounds, first in a weight compressed network and then in an incrementally bitwise version with the same topology. The pretrained weights of the first round are used to initialize the weights of the bitwise network. For the BGRU cells, we redefined the feedforward procedure with bitwise values and operations. Due to the sensitivity in training the BGRU network, the bitwise feedforward pass is performed gently using two types of masks that determine the level of sparsity and rate of binarization. With 4-bit QaD quantized input magnitude spectra and IBM targets, the BGRU at full binarization performs well for the speech denoising job with a minimal computational cost.

\vfill\pagebreak

\bibliographystyle{IEEEbib}
\bibliography{strings,main}

\end{document}